\documentclass[twocolumns]{raa_twocolumns}
\usepackage{graphicx,times,epsf}             
\usepackage{longtable}
\usepackage{natbib}
\usepackage{blindtext, rotating}

\usepackage[a4paper=true,pagebackref=true]{hyperref}
\hypersetup{colorlinks = true, linkcolor = green, anchorcolor = red, citecolor = blue, filecolor = red, pagecolor = red, urlcolor = red}
\begin{document}
\title{Detecting and Monitoring Tidal Dissipation of Hot Jupiters in the Era of SiTian}
\volnopage{Vol.0 (200x) No.0, 000--000}      
   \setcounter{page}{1}          
\author{
Fan Yang
\inst{1,2,3}
\and Wei Wang\footnotemark[1]
\inst{2,3}
\and Xing Wei\footnotemark[2]
\inst{1}
\and Hui Zhang\footnotemark[3]
\inst{4}
\and Ji-Lin Zhou
\inst{4}
\and Su-Su Shan
\inst{2,3}
\and Jie Zheng
\inst{2}
\and Wei-Kai Zong
\inst{1}
\and Ming Yang
\inst{4}
\and Yu Bai
\inst{2}
\and Song Wang
\inst{2}
\and Jia-Chen Zheng
\inst{1}
\and Yu-Ru Xu
\inst{1}
\and Yu-Feng Li
\inst{1}
\and You-Jun Lu
\inst{2,3}
\and Ji-Feng Liu
\inst{2,3}
}

\institute{
Department of Astronomy, Beijing Normal University, Beijing 100875, China; \\
\and
Key Laboratory of Optical Astronomy, National Astronomical Observatories, Chinese Academy of Sciences, Beijing 100101, China; \\
\and
School of Astronomy and Space Science, University of Chinese Academy of Sciences, Beijing 100049, China;\\
\and
School of Astronomy and Space Science, Nanjing University, Nanjing 210023, Jiangsu, China\\
}   
\abstract{Transit Timing Variation (TTV) of hot Jupiters provides direct observational evidence of planet tidal dissipation. Detecting tidal dissipation through TTV needs high precision transit timings and long timing baselines. In this work, we predict and discuss the potential scientific contribution of SiTian Survey in detecting and analyzing exoplanet TTV.
We develop a tidal dissipation detection pipeline for SiTian Survey that aims at time-domain astronomy with 72 1-meter optical telescopes. The pipeline includes the modules of light curve deblending, transit timing obtaining, and TTV modeling. 
SiTian is capable to detect more than 25,000 exoplanets among which we expect $\sim$50 sources showing evidence of tidal dissipation. We present detection and analysis of tidal dissipating targets, based on simulated SiTian light curves of XO-3b and WASP-161b. The transit light curve modeling gives consistent results within 1$\sigma$ to input values of simulated light curves. Also, the parameter uncertainties predicted by Monte-Carlo Markov Chain are consistent with the distribution obtained from simulating and modeling the light curve 1000 times. The timing precision of SiTian observations is $\sim$ 0.5 minutes with one transit visit. We show that differences between TTV origins, e.g., tidal dissipation, apsidal precession, multiple planets, would be significant, considering the timing precision and baseline. 
The detection rate of tidal dissipating hot Jupiters would answer a crucial question of whether the planet migrates at an early formation stage or random stages due to perturbations, e.g., planet scattering, secular interaction.
SiTian identified targets would be constructive given that the sample would extend tenfold. 
}

\authorrunning{Yang et al.}            
\titlerunning{Tidal Dissipation Detection in SiTian}  
\maketitle
\footnotetext[1]{wangw@nao.cas.cn}
\footnotetext[2]{xingwei@bnu.edu.cn}
\footnotetext[3]{huizhang@nju.edu.cn}

\section{Introduction}
Tidal migration is one of mechanisms that may explain why hot Jupiters occur at such close orbits, though direct observational evidence of orbit decay is only achieved for a single planet, namely WASP-12b, in the past decades \citep{2017wasp12b, Dawson2018, 2021wasp12bTESS}. The first data release of Transiting Exoplanet Survey Satellite \citep[TESS;][]{Ricker2015} in 2018, leads to the identification of four new candidates through TTV detection \citep{Dong2021,Davoudi2021,shan2021,Yangwasp161,yangxo3b}. 
TTV monitoring depends on the timing precision and baseline length which are key technical specifications for time-domain facilities, e.g., Zwicky Transient Facility \citep[ZTF;][]{Bellm2019}, Large Synoptic Survey Telescope \citep[LSST;][]{LSST}, Tsinghua University-Ma Huateng Telescopes for Survey \citep[TMTS;][]{Zhang2020, LinJie2021}, Antarctic Survey Telescopes times 3 \citep[AST3;][]{Ma2018, Zhang2019}, and SiTian \citep{SiTian, Zhu2021}. The latter is aiming to tune down the false-positive probability of time-domain signals, by implementing a global network of three-color photometric monitoring of one-quarter of the sky at a cadence of 30min, down to a detection limit of $V\sim21$ mag \citep{SiTian}.

It may suggest a TTV signal if any constant period ephemeris can not fit well the observed timings. TTV could originate from multiple physical processes, e.g., tidal dissipation, apsidal precession, R$\o$mer effect, and mass loss \citep{Ragozzine2009, Valsecchi2015, 2017wasp12b,Ou2021}. Another TTV generation process is the interaction between planets in a multi-planet system which should cause oscillation in transit timing residual in an observable timescale \citep{Holman2010, Weiss2013}. However, it has been reported that a planet companion should not be close or massive enough to induce observable TTVs for hot Jupiters \citep{Huang2016}. Distinguishing among the above mentioned various physical origins requires long-term high precision monitoring of transit timings continuously.

Combining observations from wide field transit surveys, e.g., Kepler, TESS \citep{Kepler2010, Ricker2015} and follow-up observations using focused telescopes e.g., CHEOPS \citep{Benz}, has been proved to be feasible in TTV detecting and related researches \citep{Cheops2021}. \citet{RAATTV} report TTVs of HAT-P-12b in the baseline of ten years, using light curves from small ground-based telescopes. The transit timings of WASP-32b are monitored and no significant TTV is found with the available data \citep{Sun2015}. 
WASP-43b is reported as a candidate showing period decay \citep{Jiang2016} which arises wide attention and scientific discussions \citep{Davoudi2021,Garai2021}. WASP-4b presents a significant TTV and is furtherly explained by the R$\o$mer effect \citep{WASP-4b,wasp-4b2020}. HAT-P-25 is reported with no significant TTV which sets a limit on the possible planet-planet interactions \citep{Wang2018}.
The success in observation in turn motivates the development of planet formation and evolution theory \citep{Dawson2018, Wang2019, LiuJi}.  

The sample for tidal dissipation studies would be significantly enlarged in the next decade, especially through the joint survey of ground-based telescopes, e.g., SiTian and the next generation exoplanet space telescopes, e.g., Habitable ExoPlanet Survey \citep[HEPS;][]{YuHEPS, Yu2020}, Earth 2.0 Transit Planet Survey, and shall be studied in details by  ARIEL \citep{ARIEL}, HABItable Terrestrial planetary ATmospheric Surveyor \citep[HABITATS;][]{HABITATS}.

In this work, we discuss the potential contribution of SiTian in detecting and investigating tidal decaying hot Jupiters. 
We present general-purpose tools for tidal dissipation detection, using the simulated light curves of SiTian. 
The detection pipeline contains four major steps, i.e., light curve generation, contamination light deblending, transit timing detection, and transit timing modeling. The latter three modules are described given that light curve generation would be integrated into SiTian science processing pipeline \citep{SiTian}. The paper is organized as follows. In Section 2, we introduce the SiTian exoplanet observation strategy and the generation of simulated SiTian light curves. In Section 3, we present the tidal dissipation identification pipeline and apply it to the simulated light curves of WASP-161b and XO-3b. In Section 4, we discuss the potential contribution of SiTian in planet tidal dissipation. In Section 5, a summary is presented.

\section{SiTian Exoplanet Observational Strategy and Simulated Transit Light Curves Generation}

SiTian is an integrated network of 1-meter telescopes, aiming at time-domain astronomy \citep{SiTian}. It shall produce a huge amount of the  light curves, that would contribute to time-domain researches of targets with physical size ranging from galaxy clusters to planets \citep{lb1liu, liulb12020, Yangatmos, yangagn, yangbinary,LennonHSTLB1,WangLTD, yang2021ltd064402245919, Ngeow2021}. 
Monitoring the same area of the sky simultaneously in three bands ($u$, $g$, $i$), SiTian will deliver high precision timing measurements, enabling detections of TTVs. The commissioning of SiTian is expected to be before or around 2030, and the first three proto-type equipment may start its operation in the middle of 2022.

\subsection{SiTian Technical Specification for Exoplanet Research}

\begin{figure*}
  \centering
       \includegraphics[width=7in]{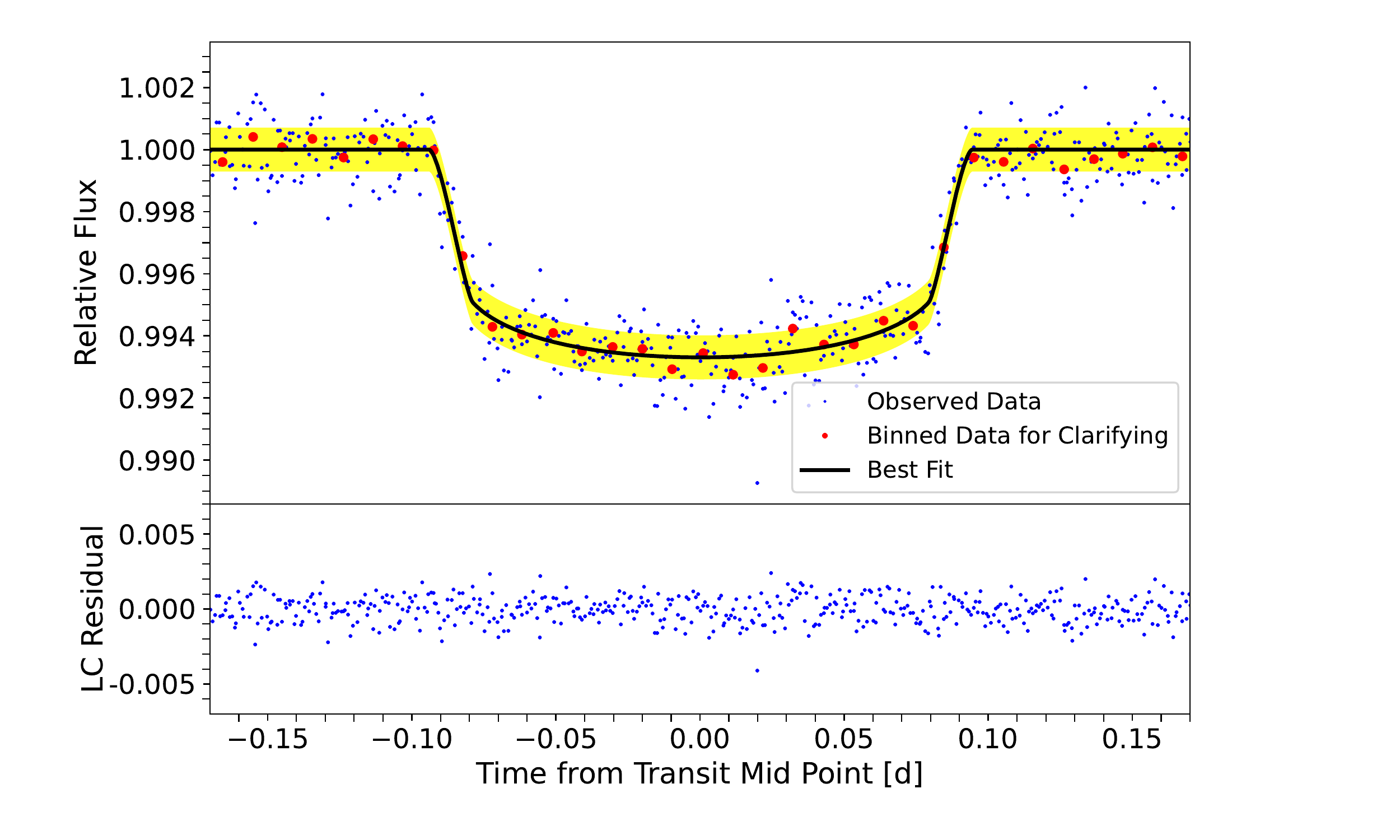}
     \includegraphics[width=7in]{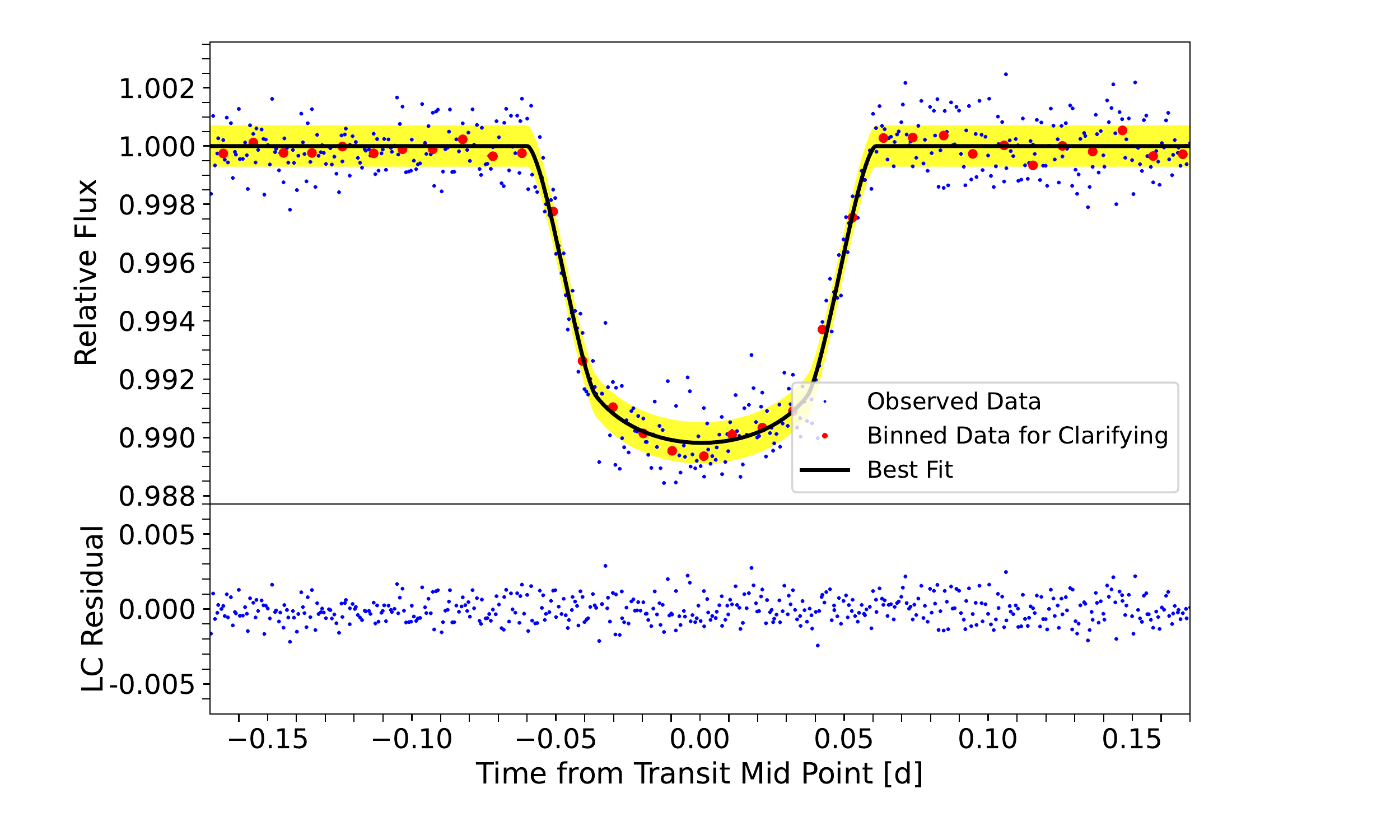}
    \caption{The simulated and model-predicted light curves. The blue points are the simulated SiTian light curve and the red points are binned data for clarity. The black line shows the best-fit model light curves, while the yellow region presents 1 $\sigma$ significance region. The fitting residuals are shown below the light curves. The top panel refers to WASP-161b and the bottom panel refers to XO-3b.
}
\label{image: lc} 
\end{figure*}

SiTian will eventually accomplish with a worldwide network of 72 telescopes, monitoring a sky area of $\sim$ 30,000 deg$^{2}$ \citep{SiTian}. The main monitoring sky coverage is $\sim$ 10,000 deg$^{2}$ which can be observed by telescopes located in China. The typical exposure time is 1 minute, resulting in a 5 $\sigma$ brightness limit of 21.1\,mag in the $g$-band \citep{SiTian}. The photometric precision is expected to reach 1$\%$ for point sources with $V = 16$ mag \citep{SiTian}, assuming the detector has a QE of 70$\%$, read noise of 7 e$^-$, an optical throughput of 70$\%$, filter transmission fraction of 70$\%$, seeing of 2.5 arcsec, and night-sky brightness of 21.1 mag arcsec$^{-2}$. The applied CMOS detector GSENSE4040 has a readout noise of 3.7 e$^-$ \citep{SiTian}. A classic good observing site has an average night-sky brightness around or better than 22.0 mag arcsec$^{-2}$ and a median seeing size better than 1 arcsec, for example, the Lenghu site\citep{Deng2021Natur}, which is the most promising major Chinese site for SiTian. Using these parameters for calculation, the photometric precision is predicted as shown in Table \ref{table: precision}. The photometric precision is $\sim$ 700 parts per million (ppm) at $g$-band and 1075 ppm at $i$-band for 12 mag targets (as shown in Figure \ref{image: lc}). The photometric uncertainties of sources brighter than 12 mag are dominated by photon noise. In addition, the SiTian mission would occupy at least three 4-meter class spectroscopic telescopes for following-up observation.

\begin{table*}
\setlength{\tabcolsep}{5mm}
\begin{center}
\caption{The expected limiting magnitude and photometric precision of a 1 minute single exposure for one SiTian telescope, based on the technical specifications from \citet{SiTian}.}
\label{table: precision}
\begin{tabular}{cccc}
  \hline
 \hline
             &    $u$    &    $g$   &    $i$  \\
\noalign{\smallskip}
\hline   
\noalign{\smallskip}
Limiting magnitude$^a$  &  20.2  &   21.1 &   20.4 \\
\noalign{\smallskip}
\hline
\noalign{\smallskip}
\noalign{\smallskip}
\multicolumn{4}{c} {Photometric Precision at Certain Magnitude} \\
\hline
\noalign{\smallskip}
Magnitude & $\sigma(u)$ & $\sigma(g)$ & $\sigma(i)$\\
\noalign{\smallskip}
16 Mag  &  0.85$\%$  &   0.48$\%$   &   0.76$\%$ \\
\noalign{\smallskip}
12 Mag$^{b}$  &    1258 ppm  &  708 ppm  &  1075 ppm  \\
\noalign{\smallskip}
\hline 
\end{tabular}
\end{center}
\begin{flushleft}
Note.
(a) 5 $\sigma$ limit magnitude. 
(b) The photometric errors of brighter sources or longer exposure time are dominated by Poisson error.
\end{flushleft}
\end{table*}

The data will be reduced by the SiTian collaboration in two modes, i.e., online and offline \citep{SiTian}. The quick data reduction, light curve retrieval, and classification will be performed in the online mode immediately after observation with a typical delay of 1 minute. In addition, alerts for transient events and other interesting targets will be triggered. Full data products, e.g. time-domain images, light curves, and target catalogs would be available and released to the public after reducing by the offline pipeline.

The capability of an observation to detect transit signal in a planetary system can be represented by the signal to noise ratio (S/R) as:
\begin{equation}
{\rm S/R} = \sqrt{\frac{t_{obs}}{P}}\frac{R_p^2}{R_{\ast}^2}/\sigma_{total},\\
\end{equation}
where $t_{obs}$ is the total integration time for the source, $P$ the orbit period, $R_p$ and $R_{\ast}$ the radius of the planet and host star, respectively. $\sigma_{total}$ is the total uncertainty in the duration of transit event, taking into account Poisson photon noise, uncorrected stellar variability, and equipment noise \citep{Kepler2010}. For example, a star with $g$=12 observed by SiTian would have a $\sigma_{total}$ of 36\,ppm when the t$_{obs}\sim$6.5 hr, a typical transit duration of hot Jupiters \citep{ExoplanetArchive, Yangatmos, Yanghats5b}. A typical stellar variability level for a low variability star like Sun is $\sim$ 10 ppm on the timescale of a planet transit \citep{Jenkins2002}. This stellar variability is significantly smaller than Poisson photon noise. The $\sigma_{total}$ is $\sim239$\,ppm for a 16 mag star in the same stacked time.

The average observation time t$_{obs}$ can be estimated by multiplying the survey time and the ratio of exposure time to scanning cadence. The total field of view (FOV) of SiTian is 600 deg$^2$. The cadence for the main sky coverage (10,000 deg$^2$) is 30 minutes. Assuming 8-hour observation time per night and 300 observing nights per year, the average annual t$_{obs}$ is 3.3 days. For Neptune-sized planet (3$R_{\oplus}$) orbiting $g$=12 mag stars with orbital periods of 30 days, the average S/N  is 8.2 in one year observation, superior to the empirical exoplanet detection threshold S/R = 7.1 \citep[][and references therein]{Kepler2010,Fressin2013,Christiansen2015}. For a $g=16$ Sun-like star, the S/R for an orbiting Jupiter-sized planet is $\sim$ 16 in one-year observation. The S/R is the same for a Neptune orbiting an M-type star. A more detailed discussion of planet detecting rate shall be presented in the following work, describing the pipeline of exoplanet detecting (Yang et al. in preparation). Here, we present a simple and general calculation.

The number of planets that can be detected by a survey depends on three factors, i.e., the planet occurrence rate, star counts, and solid angle. The solid angle equals to $R_{p}$/$a$, where $R_{p}$ is planet radius and $a$ is orbital semi-major axis \citep{Kepler2010}. The planet occurrence rate is estimated to be $\sim$ 0.15 for SiTian's capability, based on the knowledge and lessons gained from the Kepler mission \citep{Howard2012, Fressin2013, Christiansen2015}.

Taking the advantage of the high spatial resolution of $\sim$ 1 arcsec, and considering the number density of bright stars, SiTian is planned to monitor the galactic plane for exoplanet research. There are about 6.5 million stars brighter than 12 mag in the galactic plane with galactic longitude $b$ between 0$^{\circ}$ and 20$^{\circ}$ \citep{Robin2003}. This leads to an expected planet detection of 25,000 if applying an average solid angle of 2.5$\%$ \citep{ExoplanetArchive}. The expected detected planet would thereby be 250 per year, assuming 1$\%$ SiTian time is allocated for bright sources. We note that the blending of nearby sources is correctable (cf. details in Section \ref{section:deblending}) for bright target transit observation \citep{Yangatmos, Yanghats5b, YangLD}. Thereby, monitoring crowded stars in the Galactic plane is an efficient strategy for SiTian to detect transit planets.

The expected number of stars with $g\sim$15-17 mag and with absolute $b$ between 20 and 40 $^{\circ}$ is $\sim$ 1 million\citep{Robin2003}. The occurrence of Jupiter-sized planets around Sun-like stars is about 0.06 \citep{Fressin2013}, which implies the detection of 1500 Jupiter size exoplanets orbiting Sun-like stars by SiTian. In addition, about 1400 Neptune size exoplanets orbiting M-type star are expected to be detected, considering the small size of M-type stars and their large abundance of $\sim$ 70$\%$ \citep{Chabrier2003}.

The detection of these 2900 exoplanets around faint stars are byproducts of SiTian's main scientific objectives. We emphasize that with only 1$\%$ observing time on bright stars, SiTian may discover additional 2500 exoplanets in ten-year observation and up to 25000 planets with 10$\%$ time.

\subsection{SiTian Data for Transit Timing Analysis and Simulated Transit Light Curves}
\label{section:deblending}

The proposed SiTian Mission would significantly extend the sample size of exoplanets. The expected multiple timing measurements of these newly discovered targets from SiTian may already allow a transit timing analysis. Meanwhile, SiTian's main survey or follow-up discretionary program of the known planets with reported timing observation and TTV evidence will provide opportunities for a full TTV study of these targets. The planet identification package will detect exoplanet candidates and provide their preliminary orbital and planetary parameters (Yang et al., in preparation). For the research of TTV, it is necessary to build a specialized package to obtain more precise transit parameters, especially the transit timings. The potential targets for the pipeline include already-known hot Jupiters and new transiting hot Jupiters detected from the SiTian survey. Below we describe our simulation of SiTian's capability on the study of TTV and tidal dissipation.

The deblending of the light curve is crucial for the transit planet survey programs with poor spatial resolution and is particularly important for studies relying on high precision transit depth measurements \citep{Yangatmos, Yanghats5b}. We note that the high spatial resolution and sampling are one of the major advantages of SiTian.

We model and remove the blending light from the unsolved sources in the vicinity of the target for the TESS image which has a pixel size of 21 arcsec \citep{Yangatmos}. Firstly, the correlation between the fraction of blended light and the distance of the contaminating star to the target is built. The step-by-step description of blending-distance correlation is available in our previous work \citep{Yangatmos}. With this relation, one can calculate the contamination fraction of every source and their sum is the total blending fraction. The calculation needs external information from the Gaia catalog \citep{Gaia2018} on the flux and position of individual stars inside the aperture of the target star.

The deblending method that we have developed delivers TESS brightness highly consistent the Gaia brightness \citep{Yangatmos}. The derived transit depths are consistent within 1$\sigma$ with those given by the TESS Pre-Search Data Conditioning \citep[PDC;][]{PDC} modules for the comparison samples \citep{Yangatmos, Yanghats5b}. The deblending package has been provided as a general-purpose software
\footnote{\url{https://github.com/sailoryf/TESS_Deblending/
}, and will be used for the deblending and correction of the SiTian light curves.}

For this purpose, we first simulate the ``observed'' SiTian $g$-band light curves of two hot Jupiters with the host star of 12 mag. The sampling interval is set as 1 minute with the input observational uncertainty of 708 ppm (as shown in Table \ref{table: precision}). The input parameters are set as the same as those of WASP-161b and XO-3b, which shows TTV evidence \citep{Yangwasp161,yangxo3b}. For each source, we simulate 1000 transit light curves. The light curves are all set with a total observation time of 8 hours. The simulation light curves are de-trended by a polynomial fit to the several-hour out-of-transit baseline data, which is shown to be a valid approach in \citep{Yangatmos, YangLD, Yanghats5b}.

WASP-161b is a hot Jupiter orbiting an F6-type star every 5.41 days \citep{discoverpaper} and is reported to bear significant TTV related to tidal dissipation \citep{Yangwasp161}. The giant planet has a mass ($M_{p}$) of 2.49$\pm$0.21$M_{J}$ and a radius ($R_{p}$) of $1.14\pm0.06R_{J}$. The host star has a mass of 1.39$\pm$0.14$M_{\odot}$, a radius of $1.71\pm0.08R_{\odot}$. Applying a combined-fit to both the radial velocity curve and the TESS 2-minute cadence light curve, \citet{Yangwasp161} has an orbital eccentricity of $e$ = 0.34$\pm$0.03, an inclination of 89.58$\pm$0.28, and a semi-major axis in stellar radii ($a/R_{\ast}$) of 6.57$\pm$0.45 \citep{Yangwasp161}. In addition, a significant TTV is reported by combining the TESS observation and the archival timing measurements.

XO-3b is an another hot Jupiter with significant TTVs \citep{shan2021,yangxo3b}. The planet-star system has an orbit period ($P_{orb}$) of 3.19 days, an $a$ of 4.95$\pm$0.18, an inclination of 84.20$\pm$0.54, and an $e$ of 0.27587$^{+0.00071}_{-0.00067}$ \citep{Winn2008, Bonomo2017, Stassun2017}. The planet has an $M_{p}$ of 11.70$\pm$0.42$M_{J}$ and an $R_{p}$ of 1.217$\pm$0.073$R_{J}$. The host star has an $M_{\ast}$ of 1.213$\pm$0.066$M_{\odot}$ and an $R_{\ast}$ of 1.377$\pm$0.083$R_{\odot}$.

When simulating the observed light curve, and performing light curve model fitting, we used the classic planet transit model assuming a Keplerian orbit from \citet{Mandel_Agol2002}. The parameters include $R_{p}/R_{\ast}$, $a/R_{\ast}$, transit mid-point $T_C$, inclination, the argument of periapsis, the time of periapse passage, the longitude of the ascending node, and quadratic limb darkening coefficients (a1 and a2). The limb darkening coefficients are inserted from limb darkening models, depending on the stellar type of the host star and observational band \citep{claret2000, claret2011, YangLD,yang2021ltd064402245919}. We set the transit midpoint as the time zero reference for comparison purposes. Examples of synthetic light curves for WASP-161b and XO-3b are as shown in Figure \ref{image: lc}. The blending light from sources nearby is negligible given that no comparable sources within 10$''$ are detected and SiTian data has a pixel resolution of 1$''$.

\begin{figure*}
  \centering
   \includegraphics[width=7in]{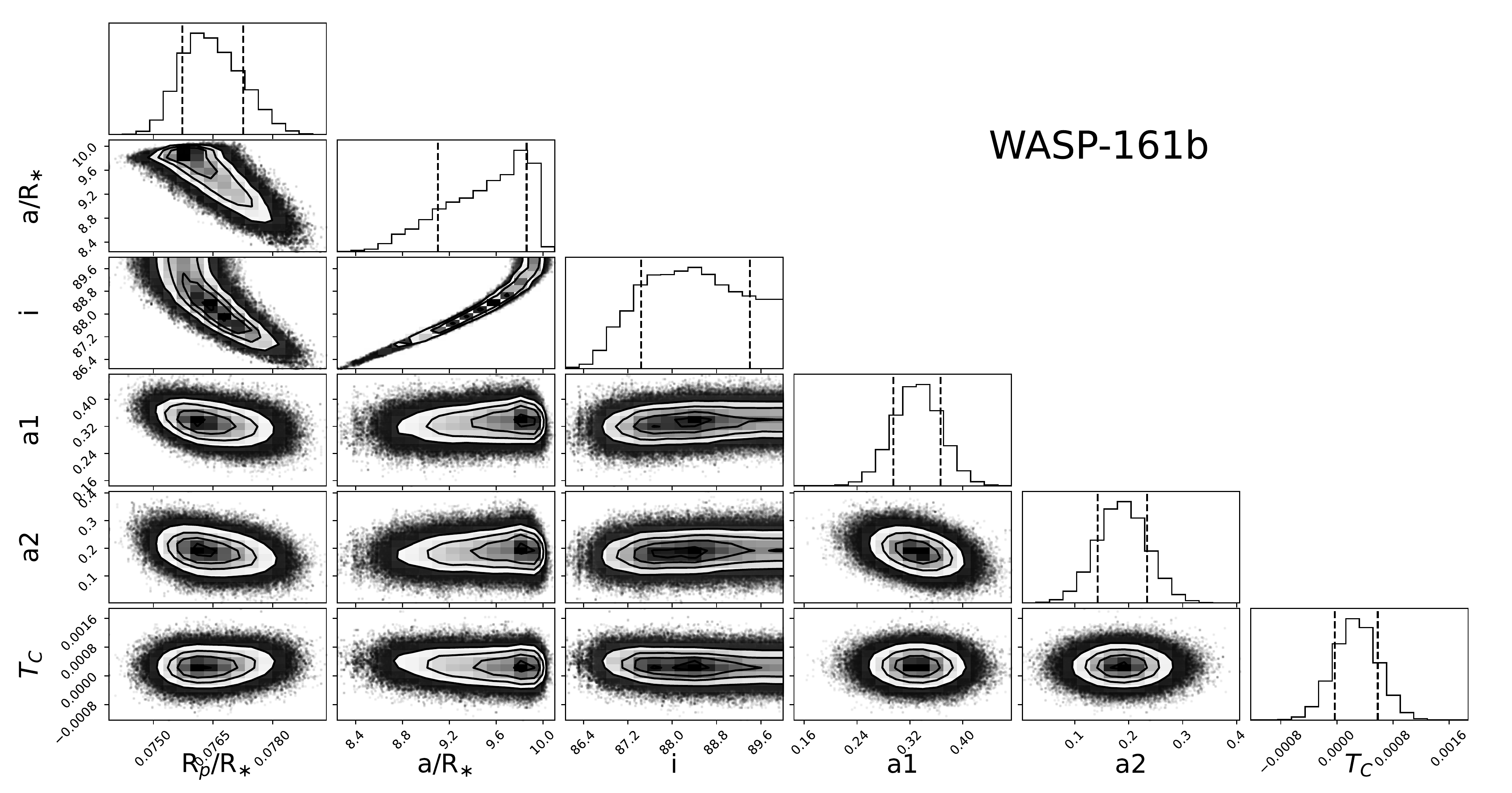}
      \includegraphics[width=7in]{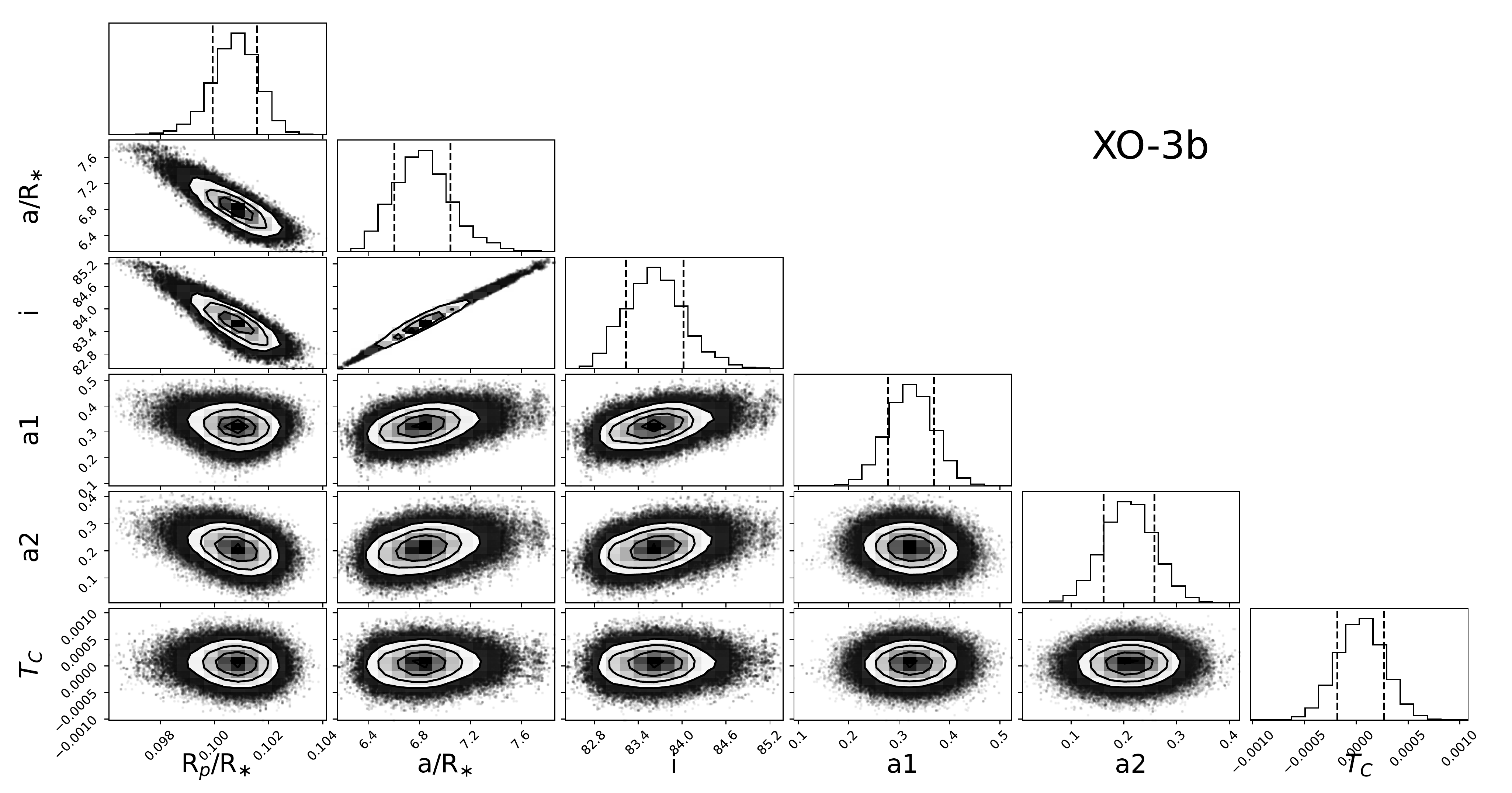}
    \caption{The posterior distribution obtained from the MCMC fitting to the simulated light curves of WASP-161b (top), XO-3b (bottom). The dashed lines in the diagonal histograms give the 1 $\sigma$ regions. 
    The corner plot uses the routine from \citet{corner2016}.
    }
\label{image: MCMC} 
\end{figure*}

\section{The pipeline for Tidal Dissipation Candidates Identification}

Our pipeline to search for the tidal dissipation candidates is firstly developed for the TESS data, including four main modules: the light curve generation and detrending, the light curve deblending from contamination by additional stars inside the aperture, transit timing determinations, and multi-epoch timing analysis. 
The light curve generation and detrending will be performed by the SiTian data reduction pipeline \citep{SiTian} which should be more specializing for SiTian data. We have described light curve deblending in Section \ref{section:deblending}.
Below, we describe timing obtaining, and timing modeling modules in detail. 
Our pipeline is based on the usage of functions from NumPy \citep{NumPy}, Astropy \citep{astropy}, EXOFAST \citep{Eastman2013}, and PYMC \citep{pymc}. NumPy and Astropy provide basic statistics and astronomical calculations. EXOFAST utilize the transit model from \citet{Mandel_Agol2002}. PYMC furnishes the Markov Chain Monte-Carlo (MCMC) technique.

\subsection{Modeling Transit Light Curve and Obtaining Transit Timing}

The Markov Chain Monte-Carlo technique \citep{pymc, EMCEE}, or alternatively, the multimodal nested sampling algorithms \citep[MULTINEST;][]{Feroz2009} are widely used for the fitting to transit light curves. These two methods have been compared in previous work and the results obtained are consistent \citep{YangLD}. In our pipeline, we currently use the MCMC method and may add the MULTINEST method as an option in future versions.

The fitting procedure applies the \citet{Mandel_Agol2002} model assuming Keplerian orbit, to generate theoretical transit light curves. All parameters involved in light curve generation are set as free parameters. Uniform priors ranging through the whole reasonable space are used for all these parameters, except for the limb darkening coefficients, which use Gaussian priors with a $\sigma$ of 0.05. The central priors of the limb darkening coefficients are dependent on stellar spectral types and can be inferred via interpolation of the classic limb darkening model \citep{claret2000, claret2011, YangLD}. The spectral types of the planet host stars can be found from Gaia catalogs. The induced uncertainties in our work can be up to a few percent in radius ratios and are not significant for the timing study \citep{YangLD}.

The MCMC algorithm omits some first steps as burn-in and applies a number of iterations for probability statistics. More iteration should more likely give an effective MCMC fitting. However, the numbers of the omitted steps and the applicable steps linearly relate to the consuming time of MCMC fitting. In practice, we show that 30,000 as burn-in numbers and 50,000 as applicable steps are high enough for the TESS light curve fitting \citep{yang2021ltd064402245919, YangLD}.

An investigation has been performed for the simulated SiTian light curves, to determine the threshold of the chain steps for securing stable MCMC results. For each of WASP-161b and XO-3b, 100 light curves are generated using their reported parameters as input, respectively. We then fit the light curves and monitor the stability of the MCMC results for the 100 run. The monitoring shows that the MCMC fitting turns stable when 10,000 steps are used as burn-in and 20,000 steps are used as analysis. Considering possible extra fluctuation due to the diversity of planet parameters and light curve quality in real applications, we apply 30,000 as burn-in and 50,000 for a statistic in the pipeline. We note that a tenfold-length chain will be used for high priority sources which already show tidal dissipation evidence, for example, WASP-161\,b, and XO-3\,b. The probability statistics of their parameters are shown in Figure~\ref{image: MCMC}.

\begin{figure}
  \centering
   \includegraphics[width=3.5in]{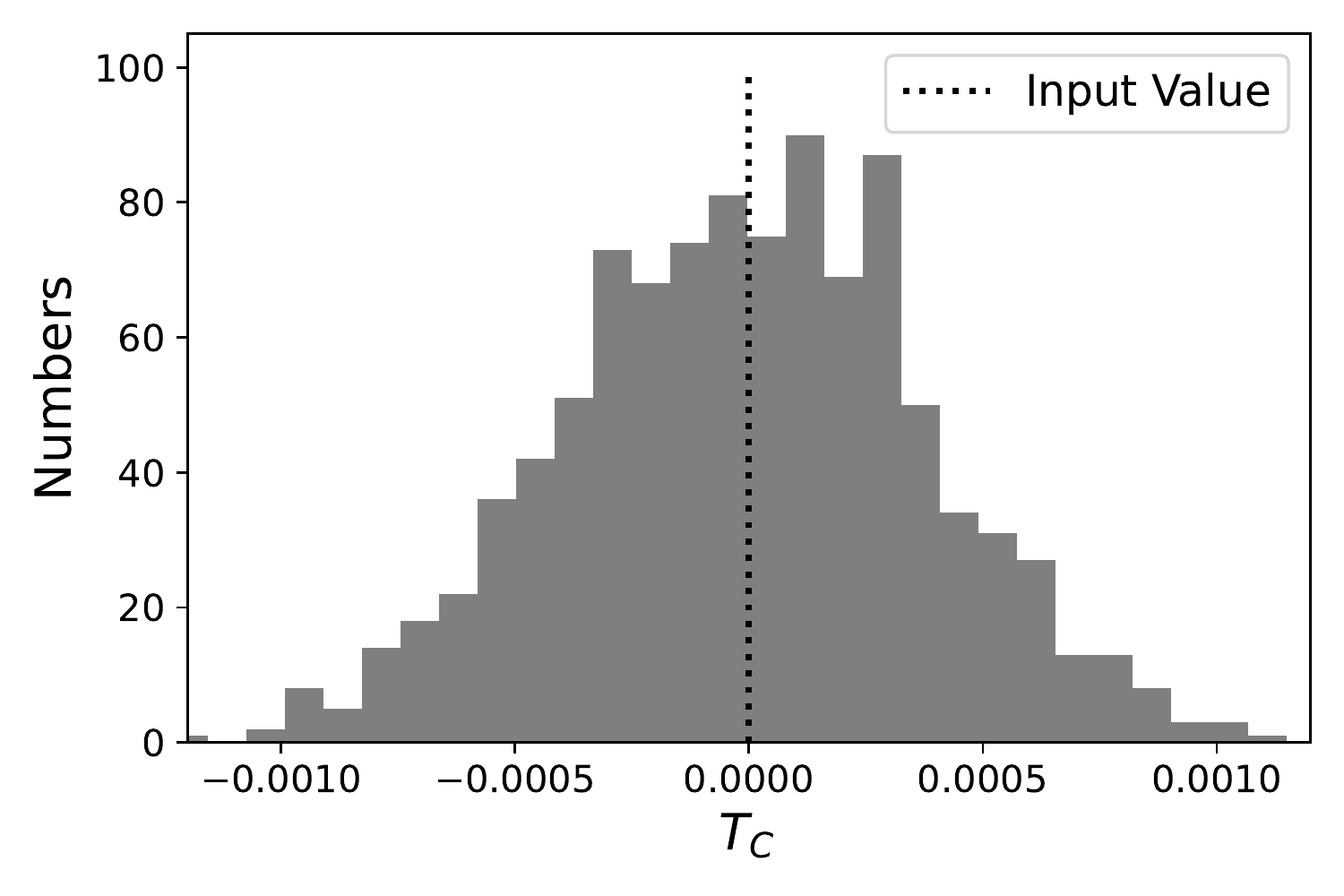}
      \includegraphics[width=3.5in]{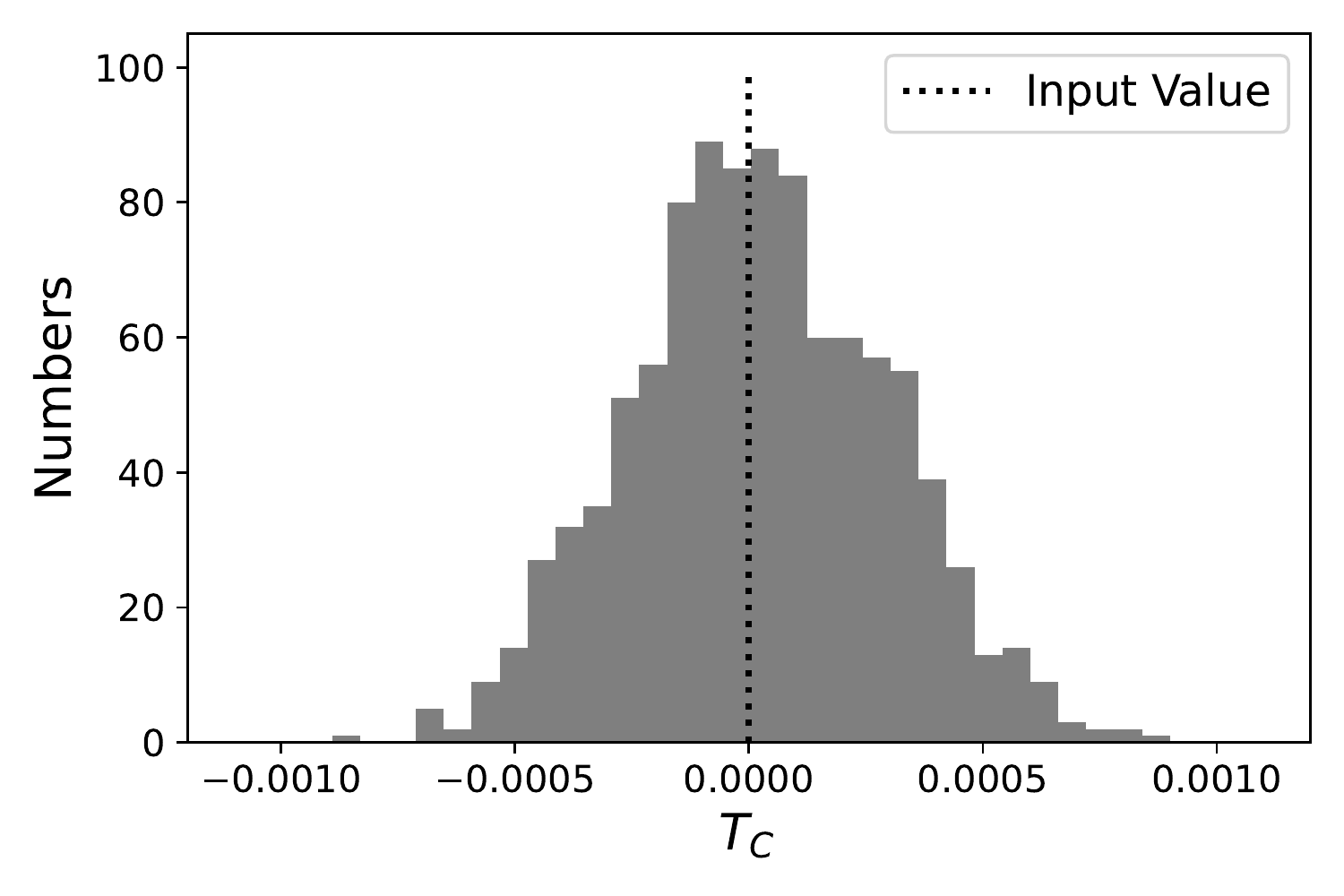}
    \caption{The distribution of the best-fit $T_C$ values of the 1000 simulated light curves. Upper panel for WASP-161\,b and lower for XO-3\,b. The vertical dash lines show the input value of 0.
    }
\label{image: his} 
\end{figure}

The MCMC fitting obtains a $T_C$ of 0.00027$\pm$0.00031 days (0.39$\pm$0.44 minutes) for WASP-161b and a $T_C$ of 0.00010$\pm$0.00022 days (0.14$\pm$0.32 minutes) for XO-3b, consistent with the input $T_C$ of 0. We apply further simulations to investigate whether the MCMC fitting is robust. We generate another 1000 light curves for each planet and fit them individually. The resulting distributions are well consistent within 1$\sigma$ to the MCMC derived distributions in fitting one particular light curve. Taking $T_C$ as example, the distributions are 0.00000$\pm$0.00038 days (0$\pm$0.54 minutes) for WASP-161b, and 0.00001$\pm$0.00027 days (0.01$\pm$0.39 minutes) for XO-3b, as shown in Figure \ref{image: his}. Therefore, we conclude that the MCMC fitting gives precise timing measurement and the timing precision obtained for the simulated SiTian light curve is $\sim$ 0.5 minutes for a single transit observation.

\subsection{Transit Timing Modeling and Tidal Dissipation Identification}

For hot Jupiters, observed TTV signal can be accounted for by various mechanisms, e.g., tidal dissipation, apsidal precession, R$\o$mer effect, multi-planet interaction, and mass loss. Exploring these physical processes is among the main scopes of SiTian for the exoplanet research. The apsidal precession scenario can be identified or rejected by studying the shape of the light curve \citep{Jordan2008, Nascimbeni2021, yangxo3b}. For the mass-loss case, the TTV signal is dependent on the mass rate and amplitude, which can be well constrained by orbital parameters. The R$\o$mer effect can be tackled by long-term radial velocity monitoring of the host star, or high-resolution imaging search for stellar-mass companion\citep{KELT-19,yang2021ltd064402245919}.

\begin{figure*}
\centering
\includegraphics[width=7in]{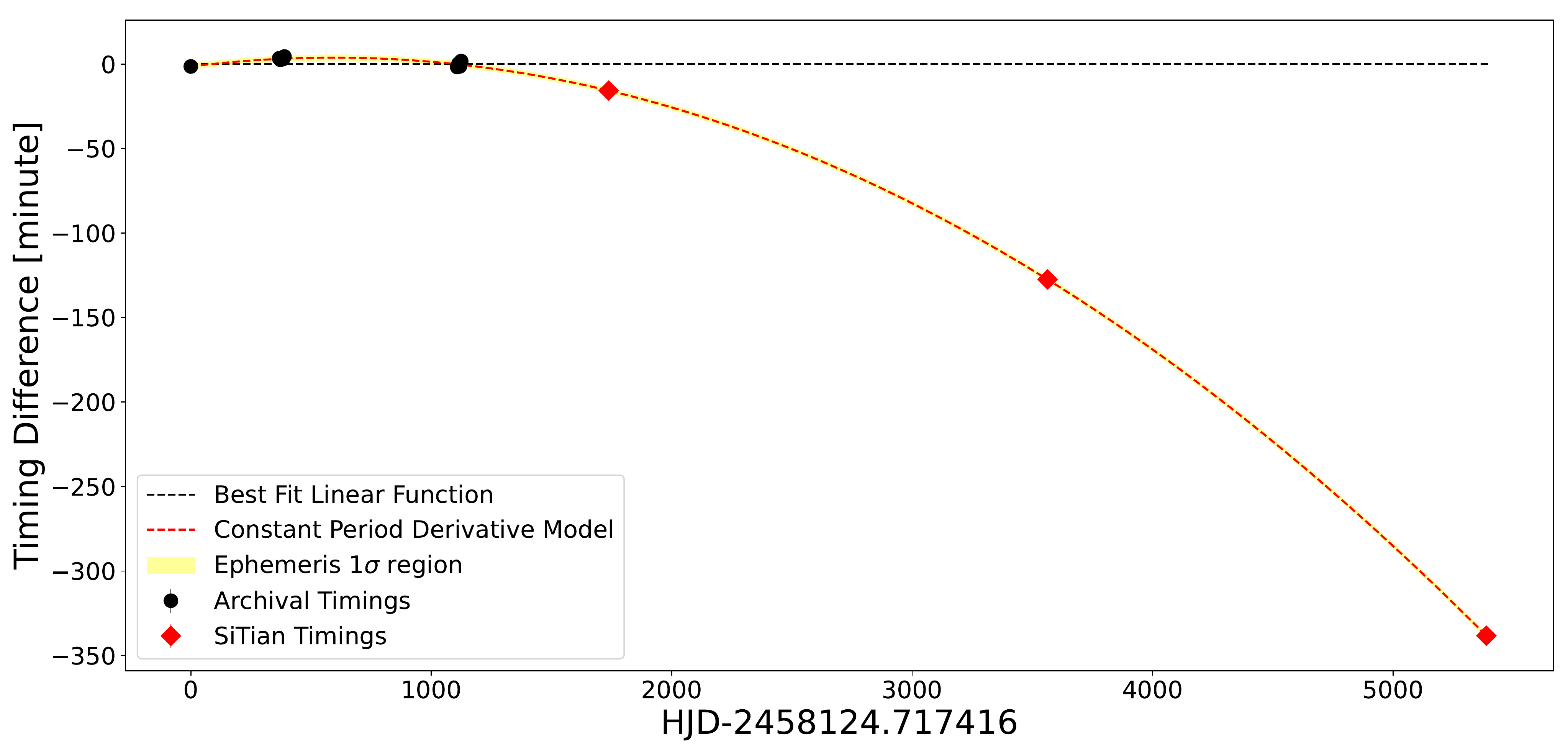}
\includegraphics[width=7in]{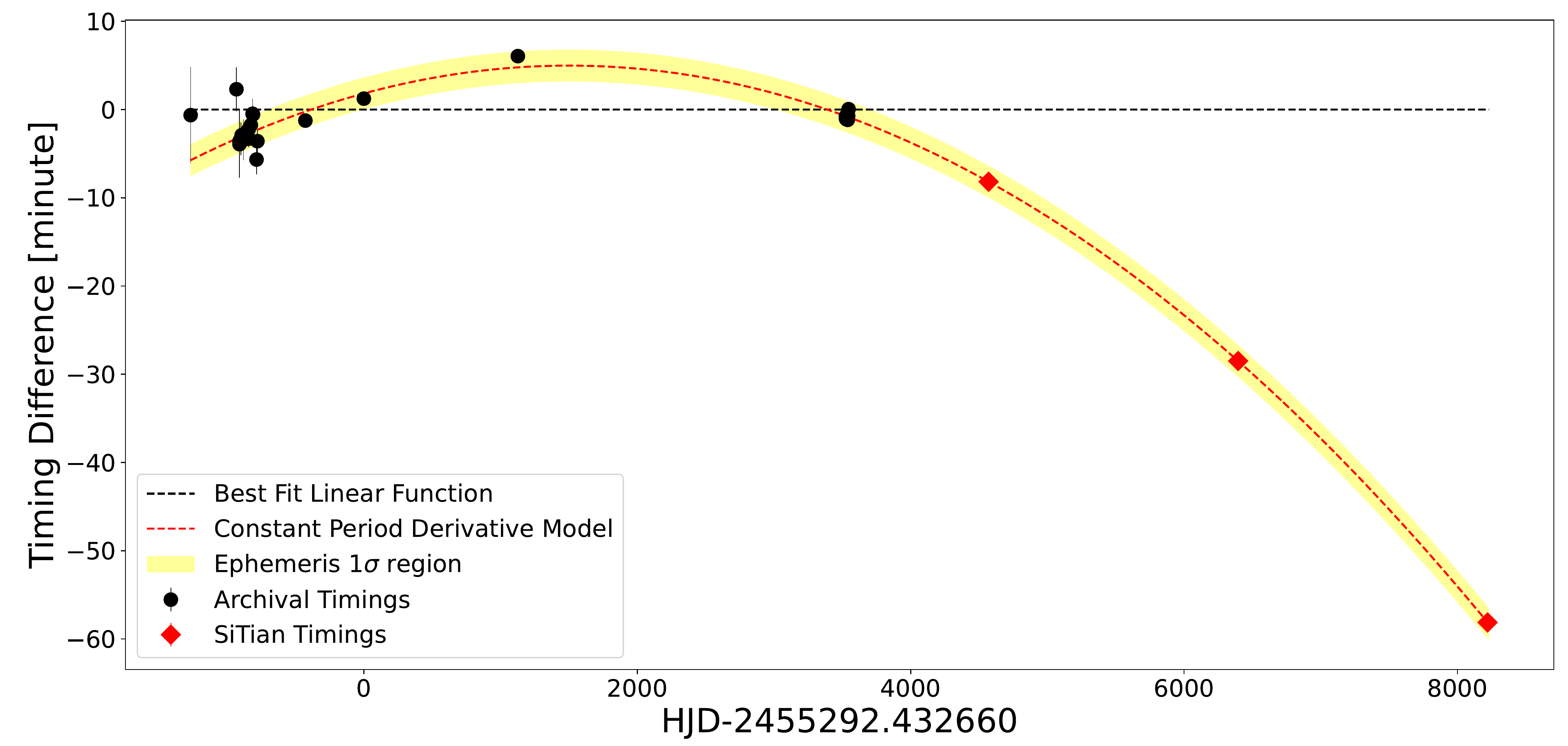}
\caption{The archival timings (black points) and the predicted new SiTian timing observations (red diamonds) of WASP-161b (top) and XO-3b (bottom). The black horizontal line gives the best fitted constant period model. The red dashed curve presents tidal dissipation models as described by \citet{Yangwasp161,yangxo3b} with the yellow region showing 1 $\sigma$ region encompassing 68$\%$ archival timings. The simulated SiTian data are to be taken in 2022, 2027, and 2032, which can effectively distinguish between the linear and the quadratic models.  
    }
\label{image: timing} 
\end{figure*}

In principle, a TTV signal induced by tidal dissipation should persist a constant period derivative, which can be easily spotted out from transit timing data using our developed pipeline (as shown in Figure \ref{image: timing}).

As predicted by the tidal dissipation model, timing variation follows a quadratic function of the transit epochs \citep{2017wasp12b, Yangwasp161}:
\begin{equation}
\label{equ: timings}
t_\mathrm{tra}(N) = t_0 + NP + \frac{1}{2}\frac{dP}{dN}N^2,\\
\end{equation}
where $t_\mathrm{tra}(N)$ is the timing of Nth transit, $t_0$ is the zero point.

It was found in previous works that timings of WASP-161b and XO-3b significantly favor a quadratic function model than linear models \citep{Yangwasp161,yangxo3b}. We present the predicted SiTian timing measurements of WASP-161b and XO-3b in 2022, 2027, and 2031 (as shown in Figure \ref{image: timing}). Transit midpoints are generated according to Equation \ref{equ: timings}. We recalculate BICs to estimate the evidence among different models. The $\Delta$BIC of WASP-161b and XO-3b is 8.8, 383 obtained from archival timings, favoring a quadratic model compared to a linear model. The $\Delta$BIC of WASP-161b and XO-3b between linear and quadratic models would be larger than 5.5$\times$10$^5$ and 1.8$\times$10$^4$, favoring quadratic models.

Adding the future SiTian measurements is shown to be highly capable of distinguishing the difference between the constant period model and the period decaying model. In our tidal dissipation detection module, a target with the BIC difference between the best fit quadratic model and the best fit linear model larger than 5 will be flagged as a candidate planet with tidal dissipation. These candidates shall be further studied and verified by follow-up observations including more transit epochs measuring and radial velocity monitoring.

We present timings and following-up strategy of KELT-19Ab as an example. KELT-19Ab is reported to hold a maximum stellar acceleration of 4 m s$^{-1}$ yr$^{-1}$ caused by the binarity of its host star \citep{KELT-19}. Applying the relation between the stellar acceleration and transit period derivative from \citet{wasp-4b2020}, the transit period would present a derivative of 5.32 ms yr$^{-1}$. \citet{shan2021} report a possible but not significant period derivative of 112$\pm$94 ms yr$^{-1}$.

\begin{figure*}
\centering
\includegraphics[width=7in]{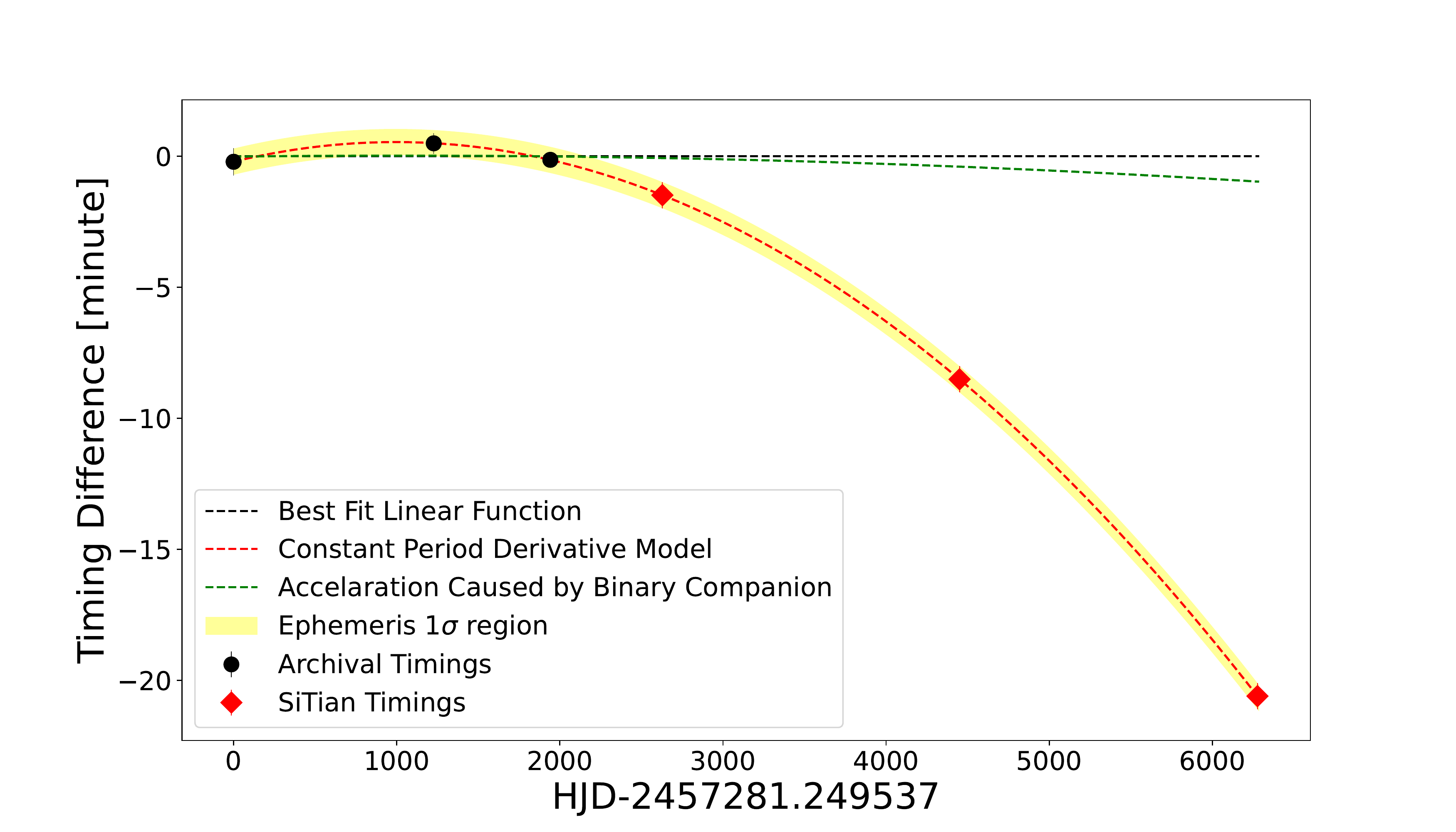}
\caption{Timings of KELT-19Ab with symbols the same to Figure \ref{image: timing}. The green line presents the period derivative caused by stellar acceleration calculated by \citet{shan2021}.}
\label{image: kelt19timing} 
\end{figure*}

Timings obtained from SiTian would potentially distinguish these different scenarios. SiTian timing predictions are obtained following the same process as described above with planet parameters taken from \citep{KELT-19}.
The timing would present a difference at $\sim$ 1.5 minutes in 2022 between a linear and a quadratic model obtained from archival timings available (as shown in Figure \ref{image: kelt19timing}). The difference would be significant when one has timings in 2027 and 2031. The $\Delta$BIC obtained from archival timings between linear and quadratic models is 2.1, slightly favoring a quadratic model. It would derive a $\Delta$BIC of 1963 until 2031 if the observational timings are as predicted by the quadratic model (seen in Figure \ref{image: kelt19timing}). Moreover, the stellar acceleration would be detectable with SiTian observation in 2031. It would reveal a 1-minute difference between the constant period and the stellar acceleration models. The period derivative due to acceleration is taken from \citet{shan2021}.

Moreover, long-term transit observation can also help to discriminate between possible TTV origins. Planets in a multi-planet system are usually not massive and close enough to each other to cause TTVs with an amplitude larger than 10 minutes \citep{Holman2010, Huang2016}. Therefore, it may rule out the possibility of the multi-planet origin, if either a monotonic TTV or a large TTV amplitude is observed, as shown in Figure \ref{image: timing}. For another important arguing possible origin, apsidal precession, the transit duration variation (TDV) should be more significant than TTV \citep{Pal2008, Ragozzine2009}. The TDV can be obtained by modeling the transit light curves as well. No evidence of TDV in the archival data is found for WASP-161b and XO-3b \citep{Yangwasp161,yangxo3b}. We expect that the SiTian project may provide additional observation on TDV.

\section{discussion}

The observed TTV signal and its interpretation are highly dependent on the data quality of timing observation. Any unexpected factors in deriving the transit timing parameters may change the period derivative \citep[detailed example in][]{Yangwasp161, shan2021}, and thus leads to different explanations. Therefore, long-baseline transit monitoring with high precision is crucial for enhancing understanding of the tidal dissipation process.

We have introduced the capacity of SiTian in modeling the tidal dissipation of WASP-161b and XO-3b (as shown in Figure \ref{image: timing}). For them, we have scheduled multi-epoch transit observations from 2022 with the SiTian prototype telescopes. In addition, continuous observation of at least one transit per five years is proposed. This observation would improve the accuracy and robustness of the period derivative, which is important for modeling the dissipation process for these two benchmark planets.

The migration is believed to be the reason that giant gas planets can exist at such en-close orbit \citep{Dawson2018}. The migration is probably due to one of the two scenarios: tidal migration that loses angular momentum by e.g., planet scattering, stellar companion scattering, and secular interaction \citep{Wu2011, Naoz2011}; or migration due to disk friction in the early stage of planet formation \citep{Lin1996}. A high eccentricity system showing period decaying favors the former model \citep{Yangwasp161, yangxo3b}. Also, the tidal dissipation occurrence rate should be higher for the former model. It is likely that both migration mechanisms may contribute separately to different sources, or even act in the same source while in different evolutionary stages.

Interestingly, the system parameters of the five reported tidal dissipation exoplanets show obvious diversity \citep{2017wasp12b, Dong2021, Davoudi2021,yangxo3b, Yangwasp161}. Extending the available sample crucial for further statistic investigations. Using the methods introduced by \citet{Dong2018, ZhuDong}, we are performing statistic analysis to find possible relationships between the eccentricity, orbital radius or semi-major axis, obliquity, metallicity, and planet-multiplicity with no conclusive results yet due to the limited sample size.

A large number of candidate systems with evidence of tidal dissipation may be discovered or identified by SiTian, which will be of great importance for understanding tidal migration. The detection number depends on the real occurrence rate of planets with tidal dissipation, which is one of the most important questions to answer by future observations, e.g. SiTian. The current fraction of such systems is 5 out of more than 4000. We note that planets other than hot Jupiters bearing tidal dissipation should be rarer. We expect to discover $\sim$ 50 new tidal dissipation star-planet systems taking into account at least one order of magnitude larger sample size. Moreover, other targets showing TTV signals but caused by other mechanisms such as planet-planet interactions, apsidal precession, will be identified by our pipeline as byproducts.

For targets with high scientific interests, following-up observation would be promptly scheduled with the SiTian 4-meter class spectroscopic following-up telescopes \citep{SiTian}. The following-up observation would give us information of e.g., host star properties, binarity, and planet multiplicity. In addition, future advanced facilities will provide important opportunities for deep investigations. For example, the HABItable Terrestrial planetary ATmospheric Surveyor (HABITATS), which is a proposed mission for a 4-6 meter space telescope aiming at modeling the atmospheric features of exoplanets down to Earth-like planets\citep{HABITATS}. The high-precision transmission spectra from HABITATS will help us understand the atmosphere properties of tidal dissipation targets, including the composition, thermal structures, inflation, and escape. This information shall yield a comprehensive understanding of the undergoing physical processes in the tidal dissipating sources.

\section{Summary}

We describe a pipeline for tidal dissipation detection as software preparation for the ongoing SiTian survey which is expected to have the first light of the three prototype telescopes in the middle of 2022. The SiTian survey will have 72 telescopes in total with full installation till 2030 \citep{SiTian}. We have estimated SiTian's capability for detecting exoplanets, based on its technical parameters. SiTian is expected to discovery 5,000 to 25,000 new exoplanets. Assuming a similar tidal dissipation occurrence, there would be $\sim$ 50 hot Jupiters showing tidal dissipation evidence.

The pipeline for tidal dissipation detection has modules of light curve deblending, transit light curve modeling, and timing modeling for tidal dissipation detection. The light curve deblending is based on our developed algorithm for TESS light curve deblending \citep{Yangatmos}. For each target, a relationship between the contamination fraction and the distance to the target will be built, based on the brightness and position measurements for individual stars from the GAIA catalog \citep{Gaia2018}, and then corrected to obtain ``uncontaminated'' light curves for SiTian.

We set the reported period decaying candidates of WASP-161b and XO-3b as examples to describe our pipeline and the contribution of SiTian data. We simulate SiTian light curves for WASP-161b and XO-3b with a photometric precision of $\sim$ 700 ppm in 1-minute cadence, estimated from SiTian's technical parameters. The MCMC technique is applied when modeling transit light curves. We show that our transit timing measurement has a precision of 0.5 minutes for a single transit observation.

We conclude that the inclusion of SiTian data shall provide key evidence to discriminate between various TTV origin models. For WASP-161b and XO-3b, a tidal dissipation origin seems to be the most likely explanation. The proposed future SiTian measurements shall significantly improve the model confidence and robustness.

\begin{acknowledgement}

We thank SiTian Collaboration for the support in preparing this work. 
The work made use of the NASA Exoplanet Archive \footnote{\url{https://exoplanetarchive.ipac.caltech.edu/index.html}} \citep{ExoplanetArchive}. We would like to thank Jin-Ping Zhu for the careful reading and constructive suggestions. We thank Bo Zhang for the helpful discussion. Fan Yang and Su-Su Shan are supported by funding from the Cultivation Project for LAMOST Scientific Payoff and Research Achievement of CAMS-CAS. Xing Wei acknowledges 
National Natural Science Foundation of China (NSFC; No.11872246, 12041301), and the Beijing Natural Science Foundation (No. 1202015). Wei Wang is supported by the National Natural Science Foundation of China (NSFC) grants No.~11988101, 42075123, the National Key RD Program of China No. 2019YFA0405102, and the science research grants from the China Manned Space Project with NO. CMS-CSST-2021-B12.
\end{acknowledgement}

\newpage
\bibliographystyle{raa}
\bibliography{ref}
\end{document}